\documentclass[prl,
 reprint,
 amsmath,amssymb,
 aps,
]{revtex4-1}

\usepackage{graphicx}
\usepackage{dcolumn}
\usepackage{bm}
\usepackage{hyperref}

\usepackage{braket}

\begin{document}

\preprint{APS/123-QE}

\title{Su-Schrieffer-Hegger (SSH) model with non-orientable bulk: Union of topology and flat bands in one dimension }

\author{Bharathiganesh Devanarayanan}%
 \email{dbharathiganesh@gmail.com}
\affiliation{%
 Physical Research Laboratory, Navrangpura, Ahmedabad, Gujarat, India -380009
}%
\affiliation{Indian Institute of Technology, Gandhinagar, Palaj, Gujarat, India - 382355.}

\date{January 14 2023 }

\begin{abstract}
In this letter we summarise our study of Su-Schrieffer-Hegger(SSH) model with a one dimensional non-orientable manifold as bulk. For this purpose a SSH model with any bulk (SAB) is introduced. We observe the following: (1) The topology of the SSH model is preserved even with a non-orientable bulk, (2) The appearance of a doubly degenerate flat band exactly at the Fermi level, (3) At half filling the band structure is metallic due to the complete overlap of valence and conduction band but the electrons will be localised because the bands are flat and (4) This system is unique because it has flat bands and at the same time is topologically non-trivial.
\end{abstract}

\maketitle





The Su-Schrieffer-Hegger(SSH) model is a model first introduced for the study of poly-acetylene with alternate strong and weak bonds \cite{sshoriginal}. Replacing strong and weak bonds by intracell and intercell hopping, this became the most pedagogic toy model for studying one dimensional topological systems by tuning the intracell and intercell hopping amplitudes\cite{KANE20133,Asboth2016}. For the sake of clarity in this letter we denote this well known SSH model as the conventional SSH (C-SSH) model. The unit cell in the C-SSH model is two sites placed linearly. There have been several generalisations of the C-SSH model like the generalised SSH model\cite{gssh1,gssh2,gssh3}, super SSH model\cite{sssh1,sssh2,sssh3}, Cruetz ladder\cite{Zurita2021tunablezeromodes}, with long range hopping\cite{lssh1,lssh2}, with a non-Hermitian Hamiltonian\cite{PhysRevA.97.052115} etc. 

However the study of topological systems over non-orientable manifolds is very scarce\cite{PhysRevB.84.193106} and there has never been an attempt at studying topological systems with a non-orientable bulk.  We introduce a framework to study such systems in one dimension called the SSH model with any bulk (SAB), as the prescription of the normal SSH model is inadequate for such an analysis. We then present our results for a particular non-orientable bulk (Note: the word Orientablity needs to be clearly defined to avoid any possible confusion. In Mathematics, it refers to the property of space that allows a consistent definition of a co-ordinate system. Here in the context of SSH model, the system is orientable if we can define neighbouring sites as belonging to same or different unit cells consistently when periodic boundary condition is applied to the bulk. The manifold that forms the bulk in our present study is non-orientable by the mathematical definition and consequently is also not-orientable in the definition limited to the SSH model. But one should note that there can be manifolds which are orientable by the mathematical definition but the chain formed by them is non-orientable in terms of the SSH model definition). 

\begin{figure}[h]
    \centering
    \includegraphics[scale = 0.125]{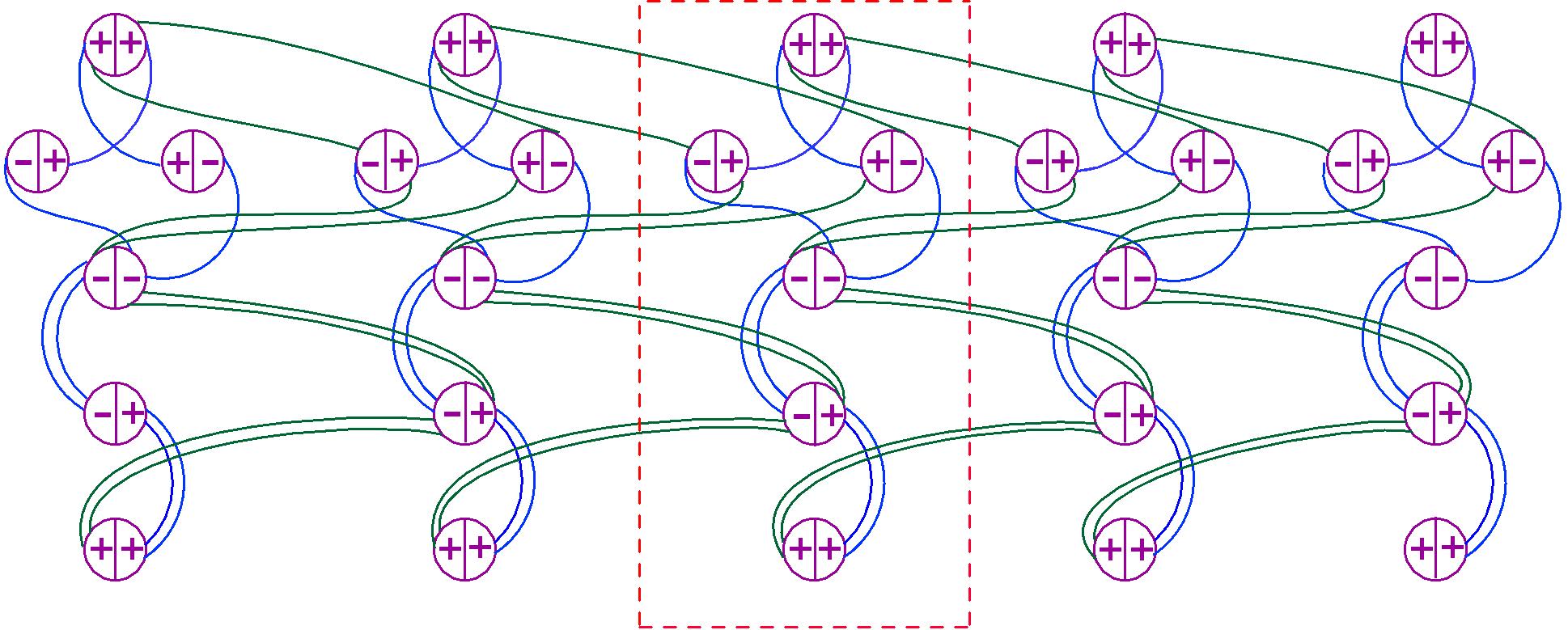}
    \caption{SAB model with a non-orientable bulk}
    \label{sshnor}
\end{figure}

\begin{figure}
    \centering\hspace{1.25 cm}
    \includegraphics[scale=0.12]{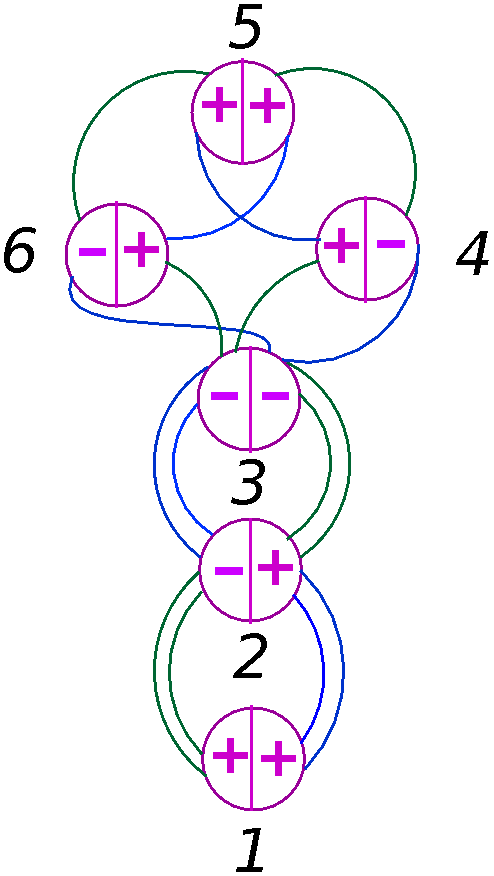}\hspace{7 mm}
    \includegraphics[scale=0.3]{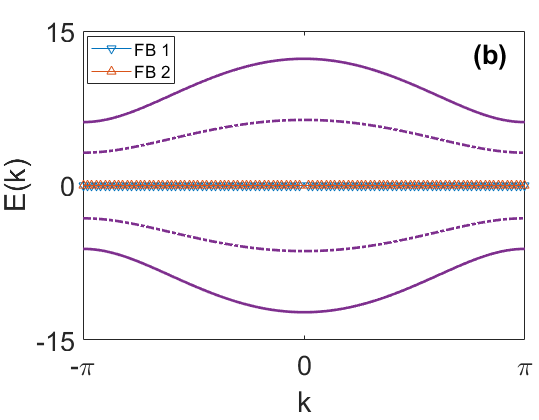}
    \includegraphics[scale=0.3]{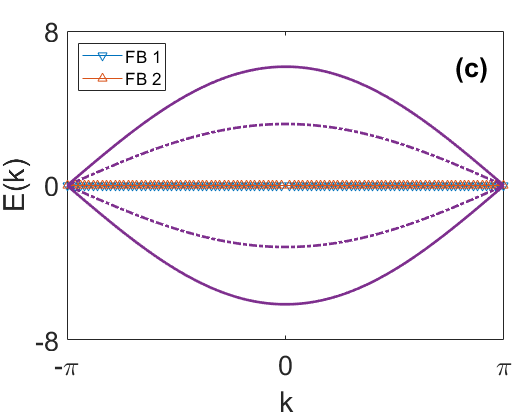}
    \includegraphics[scale=0.3]{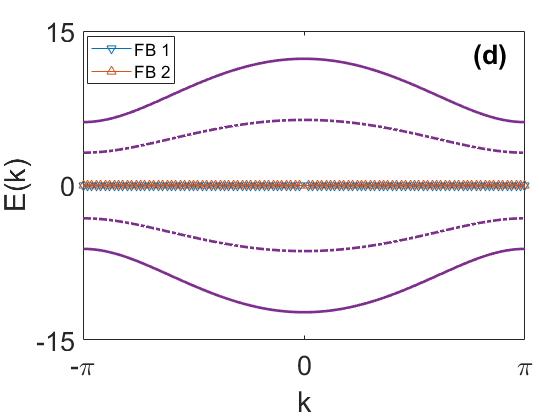}
    \caption{(a) Unit cell of SAB model with a non-orientable bulk with periodic boundary condition , band structures for the cases (b) $u>v(u = 3,v=1)$,(c) $u=v(u = 1,v=1)$, (d) $u<v(u = 1,v=3)$.}
    \label{nonor-bandstruct}
\end{figure}

We introduce here the SAB model briefly but with enough details required for this letter. The complete introduction of the SAB model along with results of the bulk replaced by several non-trivial manifolds will be given in a detailed article later\cite{sab}. In the SAB model there are two more degrees of freedom associated with each site. One degree of freedom is either left or right polarisation\footnote{The word polarisation here refers to division in to two groups and has no connection with polarisation in the context of transverse waves. Left and right are indices and have no connection with directions in real space} and each site should have one and only one left and one right polarisation. Each of this (left or right) polarisation has any one component from the other degree of freedom which has three components ($+,-,0$)\footnote{The introduction of these degrees of freedom is necessary to introduce twists in hopping so that one can realise non-orientable systems}. Let  us define the rules of the SAB model clearly:
\begin{itemize}
    \item Hopping amplitudes are non zero only when hopping is from left to left or right to right polarisation between two sites unless there is a twist. If there is a twist hopping amplitudes are non zero only if the hopping is from left to right or from right to left polarisation.
    \item Intracell hopping happens only if the two components between the two polarisations have the same components other than $0$. If the component is $0$, then are no hopping to and from that polarisation in that site.
    \item Intercell hopping happens only if the two components between the two polarisations have different components other than $0$. If the component is $0$, then are no hopping to and from that polarisation in that site.
    \item Only same cell (denoted by $u$) and nearest cell hopping (denoted by $v$) is considered.
    \item Hopping amplitudes between two sites is the same for both directions (i.e. the Hamiltonian is Hermitian).
\end{itemize}
\begin{figure}[h]
    \centering
    \includegraphics[scale = 0.5]{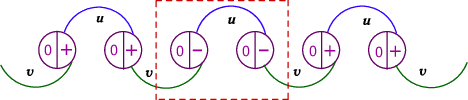}
    \caption{C-SSH model as a particular case in SAB model.}
    \label{normalsshmodel}
\end{figure}
The normal SSH model is obtained as a particular case of the SAB model in fig.\ref{normalsshmodel} (there are other similar and different varieties that appear in the SAB model\cite{sab}).
Replacing the bulk with a one dimensional non-orientable manifold requires at least six sites per unit cell and should have a twist. This is pictorially depicted in fig.\ref{sshnor}.   
The case of a line being the bulk, when considering periodic boundary condition (PBC) gives a circle, but it is not the case for the non-orientable manifold that we are considering. This is pictorially depicted in fig.\ref{nonor-bandstruct}(a). 
We will present the results for the non-orientable case when the bulk is infinite (or equivalently with periodic boundary condition) and contrast it with the C-SSH model. Firstly, the Hamiltonian in the real space for the non-orientable case for an infinite chain is:
\begin{eqnarray}
    H = \sum_{X = -\infty}^{\infty}[ 2u (c^{\dagger}_{X,1}c_{X,2} + c^{\dagger}_{X,2}c_{X,3}) \nonumber \\
    + u(c^{\dagger}_{X,3}c_{X,4} + c^{\dagger}_{X,4}c_{X,5} + \nonumber \\ c^{\dagger}_{X,5}c_{X,6} + c^{\dagger}_{X,6}c_{X,3}) \nonumber\\
    + 2v (c^{\dagger}_{X+1,2}c_{X,1} + c^{\dagger}_{X+1,2}c_{X,3}) \nonumber \\
    + v (c^{\dagger}_{X+1,4}c_{X,3} + c^{\dagger}_{X+1,6}c_{X,3} \nonumber\\ + c^{\dagger}_{X+1,4}c_{X,5}) + c^{\dagger}_{X+1,6}c_{X,5} ] + H.C.   
\end{eqnarray}
where the operator $C^{\dagger}_{X,n}(C_{X,n})$ creates (destroys) a particle at unit cell $X$ and site $n$.

\begin{figure}
    \centering
    \includegraphics[scale=0.4]{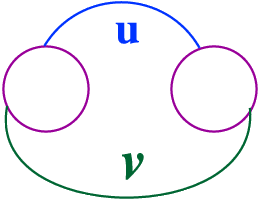}\hspace{6 mm}
    \includegraphics[scale=0.3]{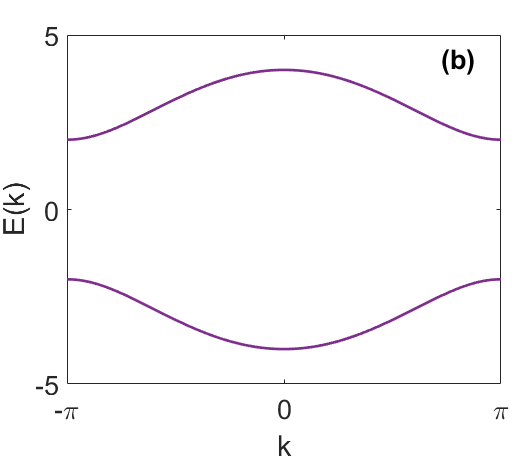}
    \includegraphics[scale=0.3]{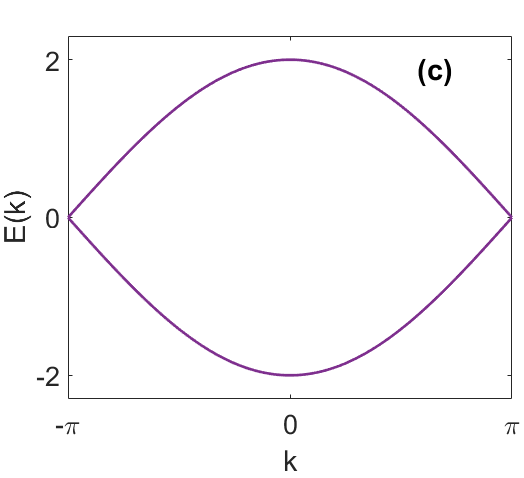}
    \includegraphics[scale=0.3]{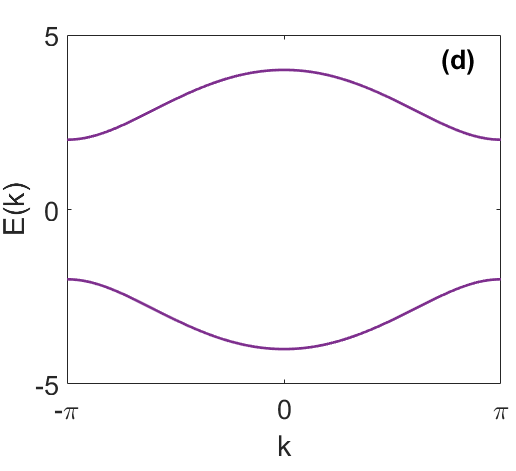}
    \caption{(a) Unit cell of C-SSH model with periodic boundary condition, band structures for the cases (b) $u>v(u = 3,v=1)$,(c) $u=v(u = 1,v=1)$, (d) $u<v(u = 1,v=3)$.}
    \label{normal SSH-bandstruct}
\end{figure}

One can write the above Hamiltonian in momentum space using Fourier transformation. We express it for the sake of clarity in the matrix form:

\begin{equation}
 H(k) = 
         \begin{bmatrix}
            0 & 2 a & 0 & 0 & 0 & 0\\
2 a^{\dagger} & 0 & 2 a^{\dagger} & 0 & 0 & 0 \\
0 & 2 a & 0 & a & 0 & a \\
0 & 0 & a^{\dagger} & 0 & a^{\dagger} & 0\\
0 & 0 & 0 & a & 0 & a\\
0 & 0 & a^{\dagger} & 0 & a^{\dagger} & 0\\
           \end{bmatrix}  
\end{equation}
where $a = u + v e^{ik}$ and $a^{\dagger} = u + v e^{-ik}$.
By diagonalising the above Hamiltonian we can get the dispersion relation of the six energy eigenvalues as a function of momentum or the band structure. The band structure for the non-orientable case is given in fig.\ref{nonor-bandstruct}. 
This is similar but different from the band structure of the C-SSH model shown in fig.\ref{normal SSH-bandstruct}. 

\begin{figure}[h]
    \centering
    \includegraphics[scale = 0.55]{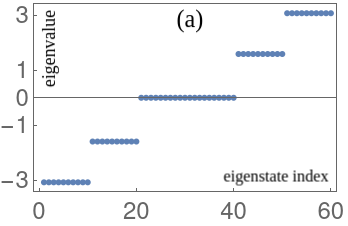}
    \includegraphics[scale = 0.55]{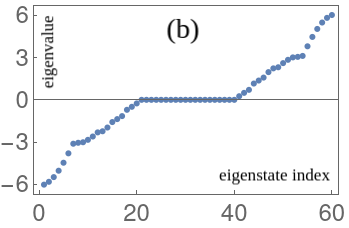}
    \includegraphics[scale =0.55]{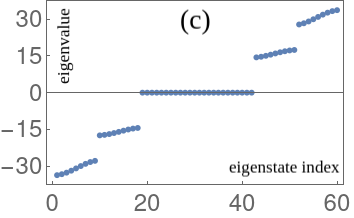}
    \includegraphics[scale = 0.22]{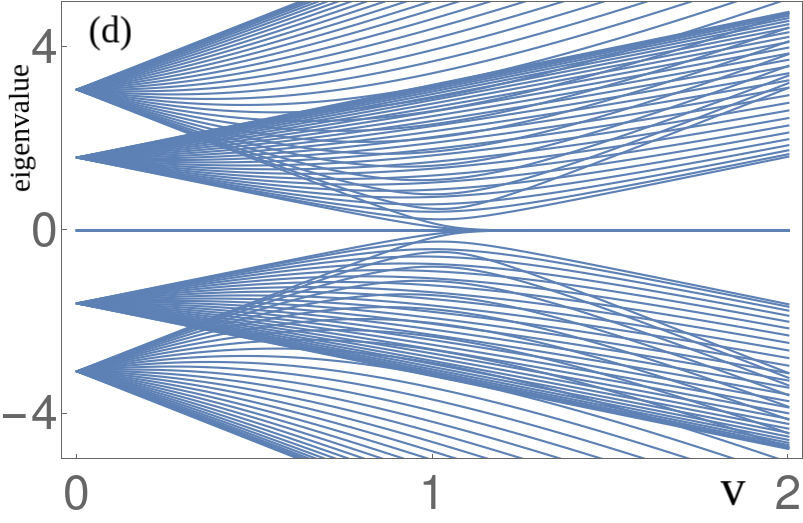}
    \caption{Energy eigenvalues for the open chain of a non-orientable bulk with 10 unit cells, sorted from low to high plotted with respect to total no. of states indexed from 1 to 60 for (a) $u>v$,(b) $u=v$, (c) $u<v$ (d) energy eigenvalues a function of intercell hopping amplitude $v$ with intracell hopping amplitude $u = 1$ }
    \label{sshnonor-open}
\end{figure}
\begin{figure}[t]
    \centering
    \includegraphics[scale = 0.5]{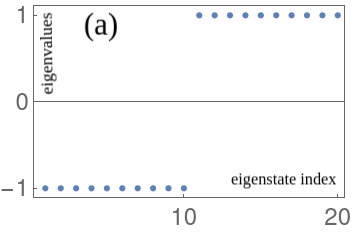}
    \includegraphics[scale = 0.55]{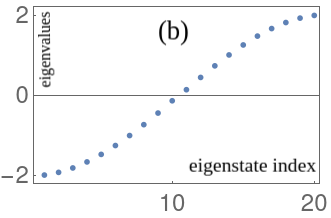}
    \includegraphics[scale =0.55]{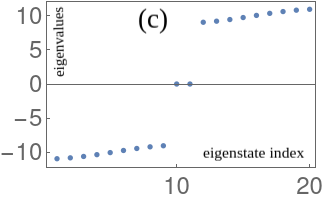}
    \includegraphics[scale = 0.25]{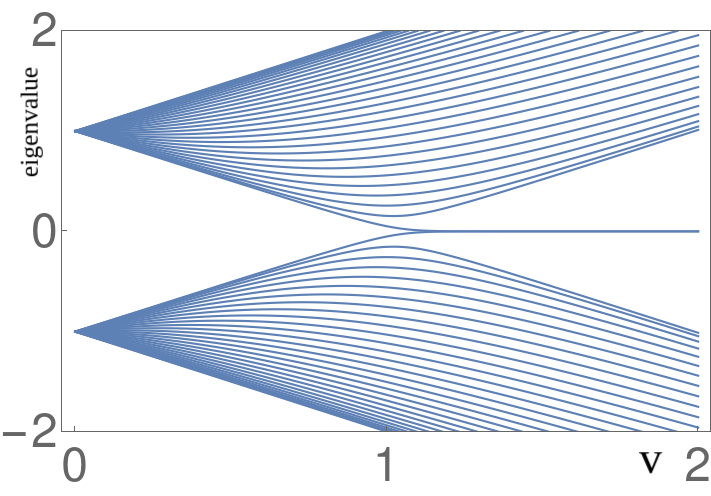}    \caption{Energy eigenvalues for the open chain of a C-SSH model with 10 unit cells, sorted from low to high plotted with respect to total no. of states indexed from 1 to 20 for (a) $u>v$,(b) $u=v$, (c) $u<v$ (d) energy eigenvalues a function of intercell hopping amplitude $v$ with intracell hopping amplitude $u = 1$ }
    \label{normalSSH-open}
\end{figure}
The following points are summarised by comparing both the band structures:
\begin{itemize}
    \item In the non-orientable bulk case, apart from the four bands, two of which are above the Fermi energy and two below the Fermi energy there is doubly degenerate flat band exactly at the Fermi energy level.
    \item This band is completely flat throughout implying a diverging value for the effective mass of electron in this band.
    \item At half filling the band structure is metallic because of the fact that the conduction and valence band completely overlap at the Fermi energy. But since the bands are flat the electronic mass at this band diverges and the electrons will be localised.
    \item If $u=v$, all the bands touch each other at the Fermi level at $k=-\pi$ and $k=\pi$ closing the band gap and indicating a topological phase transition between a trivial and topological phase as in the C-SSH model.
\end{itemize}

\begin{figure}
    \centering
    \includegraphics[scale = 0.18]{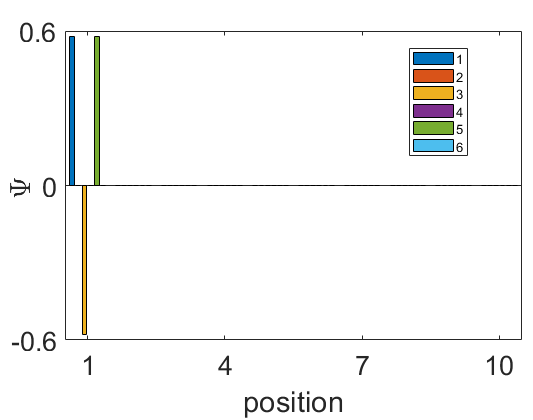}
    \includegraphics[scale = 0.18]{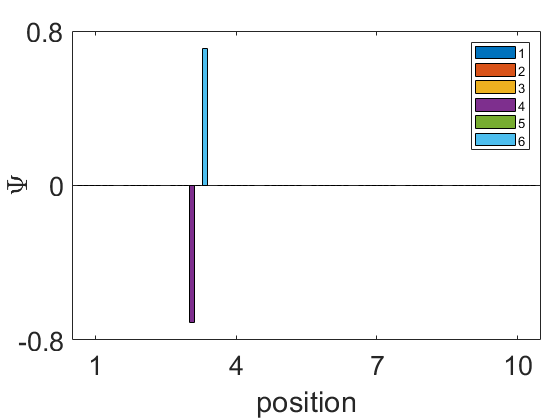}
    \includegraphics[scale = 0.18]{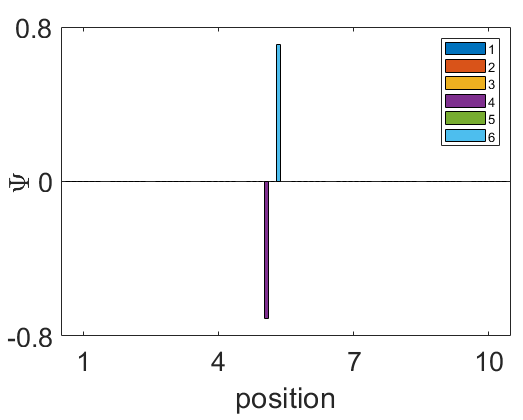}
    \includegraphics[scale = 0.19]{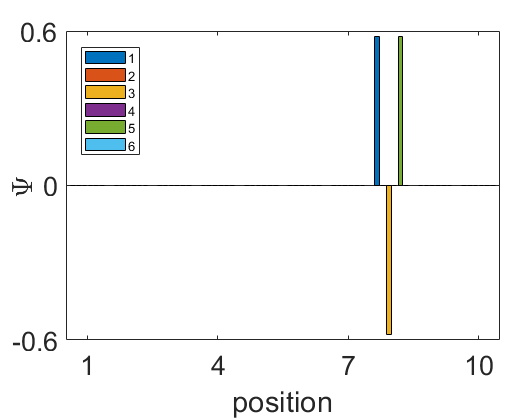}
    \includegraphics[scale = 0.19]{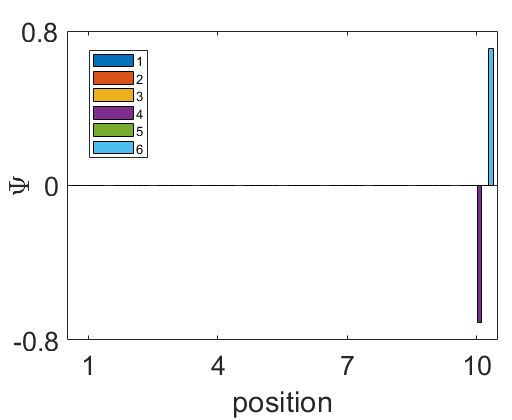}
    \includegraphics[scale = 0.19]{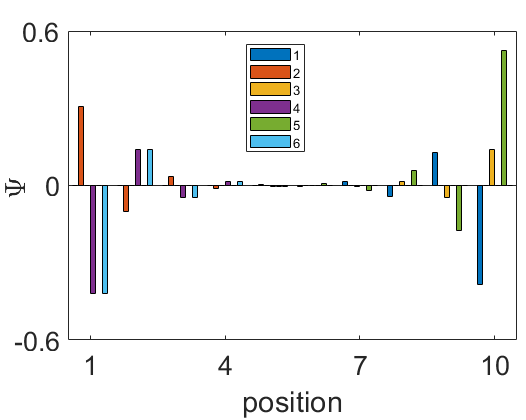}
    \includegraphics[scale = 0.19]{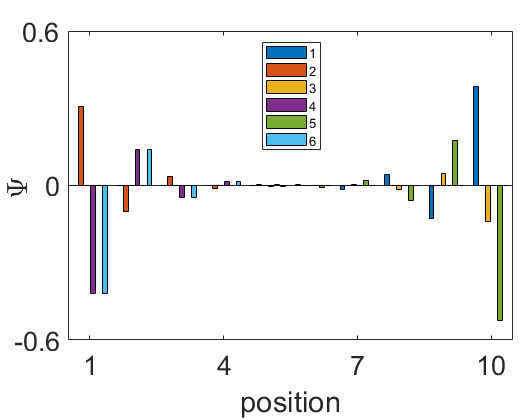}
    \includegraphics[scale = 0.19]{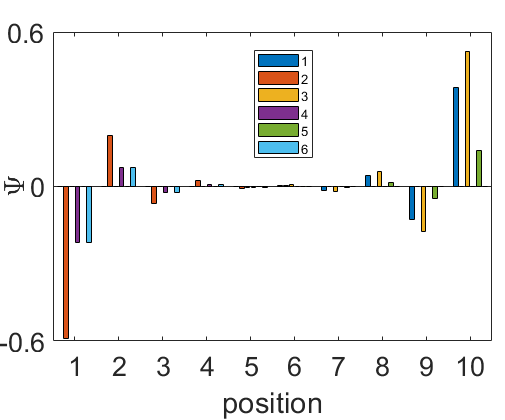}
    \includegraphics[scale = 0.19]{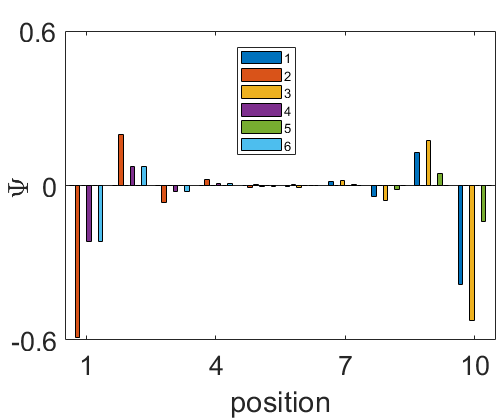}
    \caption{Wavefunctions (Normalised Eigenvectors of the Hamiltonian with 10 unit cells) corresponding to zero energy modes in the bulk (5 out of 20, (first 5)) and edge states (last 4).}
    \label{psi}
\end{figure}

We also perform the calculations with a finite bulk (or equivalently with open boundary condition). The energy eigenvalues for such a system with 10 unit cells is shown in fig.\ref{sshnonor-open}.
This is compared to the C-SSH model (fig.\ref{normalSSH-open}) and the salient points are:
\begin{itemize}
    \item In the non-orientable bulk case the trivial phase has $2N$ zero energy modes and the toplogical phase has $2N+4$ zero energy modes.
    \item It is similar to the C-SSH model because of the existence of a trivial phase when $u>v$ with no zero energy edge states and a topological phase when $u<v$ with four zero energy edge states (two in case of C-SSH model) .
    \item It is different from the C-SSH model because both the trivial and topological phases have $2N$ zero energy modes in the bulk whereas there are no zero energy modes in the bulk for the normal SSH model. This is evident from the plot of wavefunctions corresponding to various zero energies (Fig.\ref{psi}). We find that the four zero energy edge states are localised at the edges. However the states corresponding to other zero eigenvalues (that appear in both trivial and toplogical phases) are localised in the bulk.
    \item If $u=v$, then the band gaps vanishes indicating the topological phase transition point.
\end{itemize}

It has been rigorously proved in literature that for two dimensional systems, any system can simultaneously have only two out of the three features among (1) Flat bands, (2) non-zero Chern number and (3) finite hopping \cite{FBthe,flatbandtopology}. Since our system has both flat bands and finite hopping, it would be interesting to look at the presence(or absence) of the third feature. The Zak phase\cite{zak} is the one dimensional analogue of the Chern number, so we have calculated the Zak phase corresponding to the flat bands:
\begin{equation}\label{zp}
    \Phi_{n} = \int_{BZ}i\braket{u_{kn}|\partial_{k}u_{kn}}dk = \pi, \,\,\forall \,\,\,\{u,v\} \in \mathbf{R}
\end{equation}
where $u_{kn}$ is the periodic part of Bloch function corresponding to the energy band $n$, $BZ$ refers to the Brillouin zone and $n = 1,2$ correspond to the flat bands.

From Eq.(\ref{zp}), we find that the Zak phase of the flat bands is non zero ($\pi$) which is very interesting. However it remains a constant for all values of hopping amplitudes implying that the flat bands in this system do not undergo topological phase transitions\footnote{It would not be incorrect to say that the system always remains in the topological phase for which the Zak phase is $\pi$. However unlike Chern number there are certain ambiguities with the definition of Zak phase.}. 

 To summarise we had replaced the bulk in the SSH model from a linear chain to a one dimensional non-orientable manifold and investigated its consequences. Such a replacement is not possible within the framework of the conventional SSH model. For this we had introduced the SAB model of which a particular case is the conventional SSH model. We replace the bulk in the SAB model with a 1D non-orientable manifold. This system like the C-SSH model exhibits a gapped trivial and topological phase separated by band gap closing. However in both the phases there exists a doubly degenerate flat band exactly at the Fermi level. Another feature of this system is the presence of zero energy modes localised in the bulk, both in the trivial and topological phase along with zero energy edge states in the topological phase, which is in contrast to the C-SSH model where only the topological phase has zero energy modes. At half filling the band structure is metallic since the conduction and valence band completely overlap each other at the Fermi level. But since these bands are flat the electrons in these bands will be localised because of the diverging effective mass. It is also worthwhile to mention here that flat bands have been reported in various systems like the twisted bilayer Graphene\cite{PhysRevB.99.235417} and similar systems\cite{TDCs}, Kagome lattices\cite{PhysRevB.98.235109}, Lieb lattice\cite{Lieb} etc. Flat bands are associated with several exotic phases like superconductivity\cite{PhysRevB.106.104514}, Wigner crystallisation\cite{PhysRevLett.99.070401}, etc. pointing towards several potential applications\cite{aaina1,aaina2}. Here we arrive at a system with flat bands in a system with non-trivial topology in a completely different approach dealing with the orientability of the SSH model\footnote{Sri Siruvachur Madura Kali Amman Thunai, Swamiye Saranam Aiyyappa, Vetrivel Muruganuku Arogara}.     
 
 \textbf{Acknowledgment}  
 I want to thank my supervisor Prof. Navinder Singh for encouraging me to pursue this work. I also thank Dr. Paramita Dutta for her encouragement. I should also thank Ritvik, Saurabh, Supriya, Debashish and Guru for the technical helps.

\end{document}